\documentclass[aps,prl,preprint,superscriptaddress]{revtex4-1}

\usepackage{graphics}
\usepackage{subfig}
\usepackage{epsfig}
\usepackage{color}

\begin{document}


\title{First principles structures and circular dichroism spectra for the close-packed and the 7/2 motif of collagen}



\author{Karl J. Jalkanen}
\affiliation{DTU Nanotech, \O rsteds Plads Building 345E, Technical University of Denmark, DK-2800 Kongens Lyngby, Denmark\\}

\author{Kasper Olsen}
\affiliation{DTU Nanotech, \O rsteds Plads Building 345E, Technical University of Denmark, DK-2800 Kongens Lyngby, Denmark\\}

\author{Michaela Knapp-Mohammady}
\affiliation{Division of Functional Genome Analysis, German Cancer Research Center, Im Neuenheimer Feld 580, D-69120 Heidelberg, Germany\\}

\author{Jakob Bohr}
\email[]
{jabo@nanotech.dtu.dk}
\affiliation{DTU Nanotech, \O rsteds Plads Building 345E, Technical University of Denmark, DK-2800 Kongens Lyngby, Denmark\\}


\date{\today}

\begin{abstract}
The recently proposed close-packed motif for collagen is investigated using first 
principles semi-empirical wave function theory and Kohn-Sham density functional 
theory. Under these refinements the close-packed motif is shown to be stable. For the case
of the 7/2 motif a similar stability exists. The electronic circular
dichroism of the close-packed model has a significant negative bias and a large signal. An
interesting feature of the close-packed structure  is the existence of a central channel. 
Simulations show that, if hydrogen atoms are placed in the cavity, a chain of 
molecular hydrogens is formed suggesting a possible biological function for molecular hydrogen.\\

\noindent PACS\\ 87.14.em Fibrils (amyloids, collagen, etc.)\\ 87.15.ag Quantum calculations\\ 87.15.M- Spectra of biomolecules

\end{abstract}


\begin{center}
\end{center}

\maketitle


In this Letter, we consider the close-packed (CP) structure for collagen \cite{bohr2011} and subject this motif to first principles calculations and compare with calculations for the 7/2 motif. The motivation behind the CP structural motif is the remarkable coincidence that it is optimally packed while at the same time having a vanishing coupling between strain and twist \cite{olsen2012}. While the 7/2 and 10/3 structural motifs are supercoiled triple helix structures, the CP structure is akin to that of a rope where the three polypeptide strands are intertwined. 
The CP geometry is the non-periodic triple helix which optimizes the volume fraction, it has a helical pitch of 20~\AA. 

Hitherto, the 10/3 structure suggested by Ramachandran, and Rich and Crick \cite{ramachandran1954,ramachandran1955,rich1955,rich1961} and the 7/2 structure suggested by Okuyama {\it et al.} \cite{okuyama1972,okuyama1977} have been the most studied motifs for collagen \cite{beck1998} and collagen-like peptides \cite{bella1994,hongo2005}. These two motifs belong to the same symmetry class: three left-handed helical polypeptide chains are supercoiled in a right-handed arrangement. For long helical structures the relatively short range of the strand and inter-strand interactions does not favor commensuration, i.e. that an integer number of residues on the polypeptide strands corresponds to a $2\pi$ supercoiling. Hence, in general the structure will not be periodic. 
For the periodic 10/3 structure it takes ten residues to make three full $2\pi$ rotations to complete the $\sim 85$ {\AA} unit cell while for the periodic 7/2 structure it takes seven residues of one polypeptide chain to complete two full $2\pi$ rotations which complete a unit cell length of about $60$ \AA. A contributing factor to the original suggestion to change focus from the 10/3 structure to the 7/2 structure was  diffraction spots in x-ray patterns corresponding to a longitudinal period of 20 {\AA} which were unaccounted for by the 10/3 structure \cite{okuyama1999}. 

In high resolution crystallographic studies of peptide-like structures the question of whether the 7/2 or the 10/3 structure describe the data, or not, is not always resolved. R-factors indicate that both structures describe features of the diffraction data \cite{bella1994,berisio2002,okuyama2004,okuyama2006a,okuyama2006b}. Therefore atoms have been assigned differently to the three strands and hence the assignment of the atoms to the individual strands remains uncertain. And, perhaps a third structure such as the CP structure would be a better overall description which could account for some of the experimental data not yet accounted for by either the 7/2 or the 10/3 structures.  

Density functional theory (DFT) studies of various collagen-like peptides is an active research area \cite{tsai2005,palfi2008} including the interesting issue of studying biomineralization \cite{barrios2010}. For the initial configurations of the first principles calculations we used the approximate coordinates for the CP structure obtained by simple geometrical methods \cite{bohr2011} and for the 7/2 structure the coordinates obtained from the PDB depository (entry: 1K6F 3[Pro-Pro-Gly]$_{10}$) \cite{berisio2002}. These structures were then truncated to various lengths (3[PPG]$_6$, 3[PPG]$_3$ and 3[PPG]$_1$) and  hydrogens were added with the InsightII program from Biosym Technologies, Inc. To treat the effects of an aqueous environment, we have utilized the polarized continuum model (PCM) as implemented in the Gaussian 09 program \cite{gaussian09,barone98,tomasi05}. 

The semi-empirical wave function theory (WFT) PM6 model of Stewart \cite{stewart07} and two Kohn-Sham density functional theory (KS-DFT) exchange correlation functionals, the CAM-B3LYP \cite{yani04,shcherbin2008} and the LC-wPBE \cite{tawada04} have been used in this work. For the larger model,  3(PPG)$_6$, only the PM6 model was used for geometry optimization. 

Circular dichroism  is commonly used to distinguish structures with various degrees of chirality e.g. $\alpha$-helices, $\beta$-sheets, etc. The electronic circular dichroism (ECD) spectra were simulated at the CAM-B3LYP time dependent DFT (TD-DFT) level of theory for only the two smaller models, the 3(PPG)$_3$  and the 3(PPG)$_1$ systems. The 81 lowest energy singlet transition energies were calculated along with the corresponding dipole strengths and rotational strengths in both the length and velocity formalisms for both 7/2 and CP  3(PPG)$_3$ structures, and the 27 lowest singlet transition energies were calculated along with the corresponding dipole strengths and rotational strengths in both the length and velocity formalisms for both CP and 7/2 3(PPG)$_1$  structures.

\begin{figure}
\subfloat[]{
\label{fig:stacksub:a} 
\includegraphics[width=0.25\linewidth]{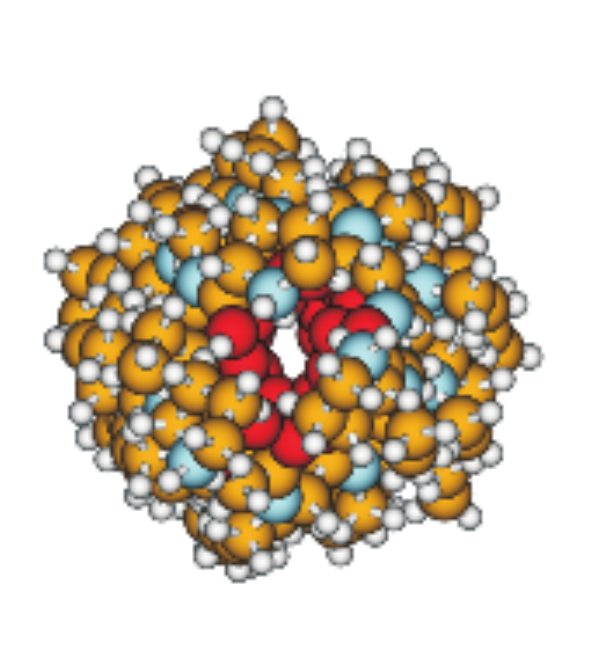}}
\hspace{0.1\linewidth}
\subfloat[]{
\label{fig:stacksub:b} 
\includegraphics[width=0.25\linewidth]{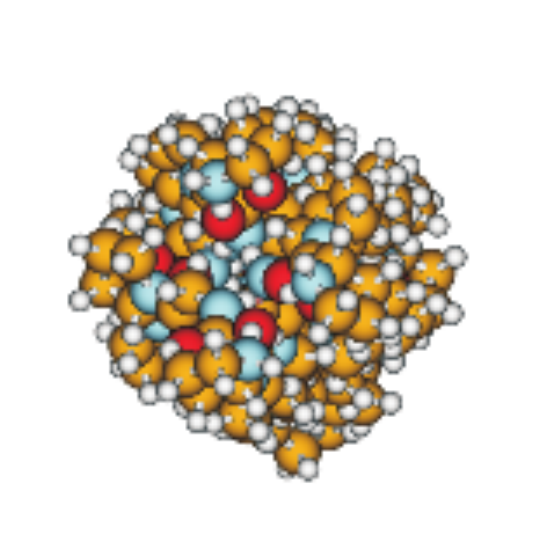}
}\\[20pt]
\subfloat[]{
\label{fig:stacksub:c} 
\includegraphics[width=0.25\linewidth]{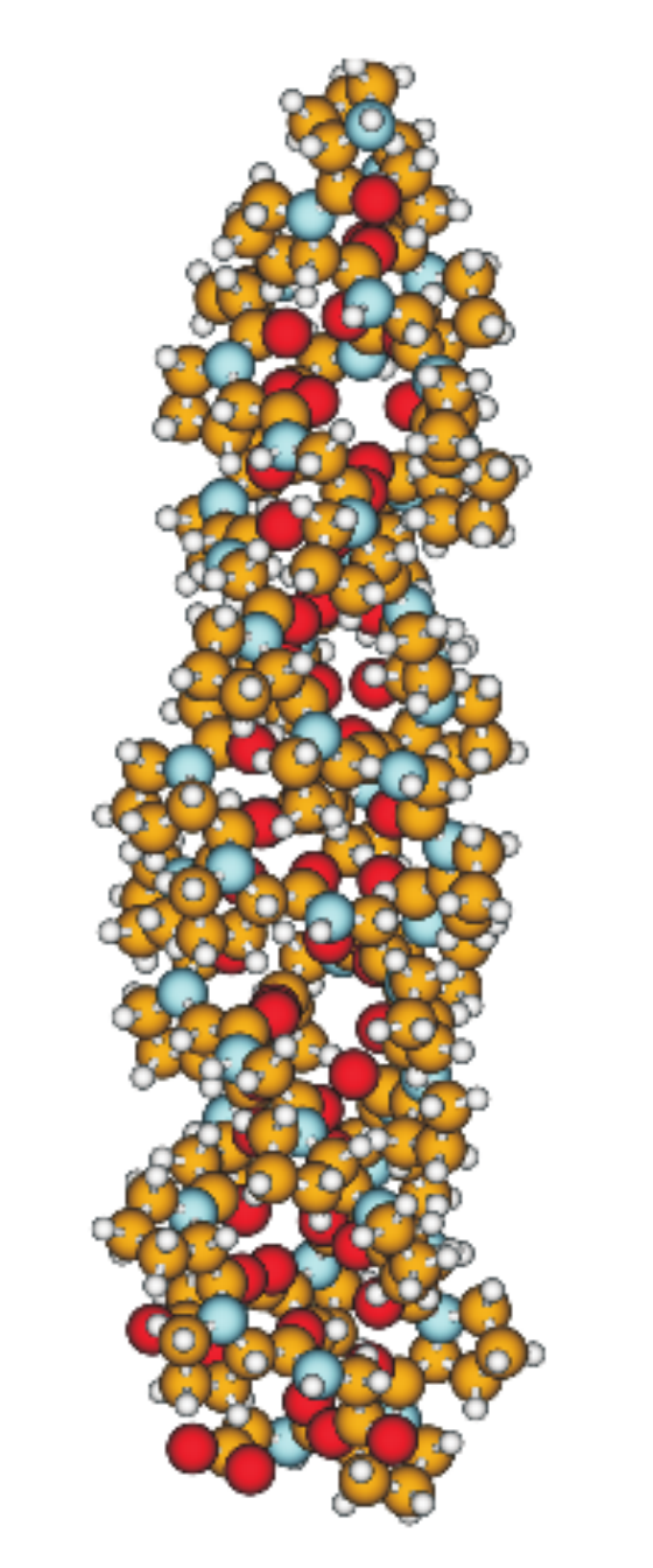}}
\hspace{0.1\linewidth}
\subfloat[]{
\label{fig:stacksub:d} 
\includegraphics[width=0.25\linewidth]{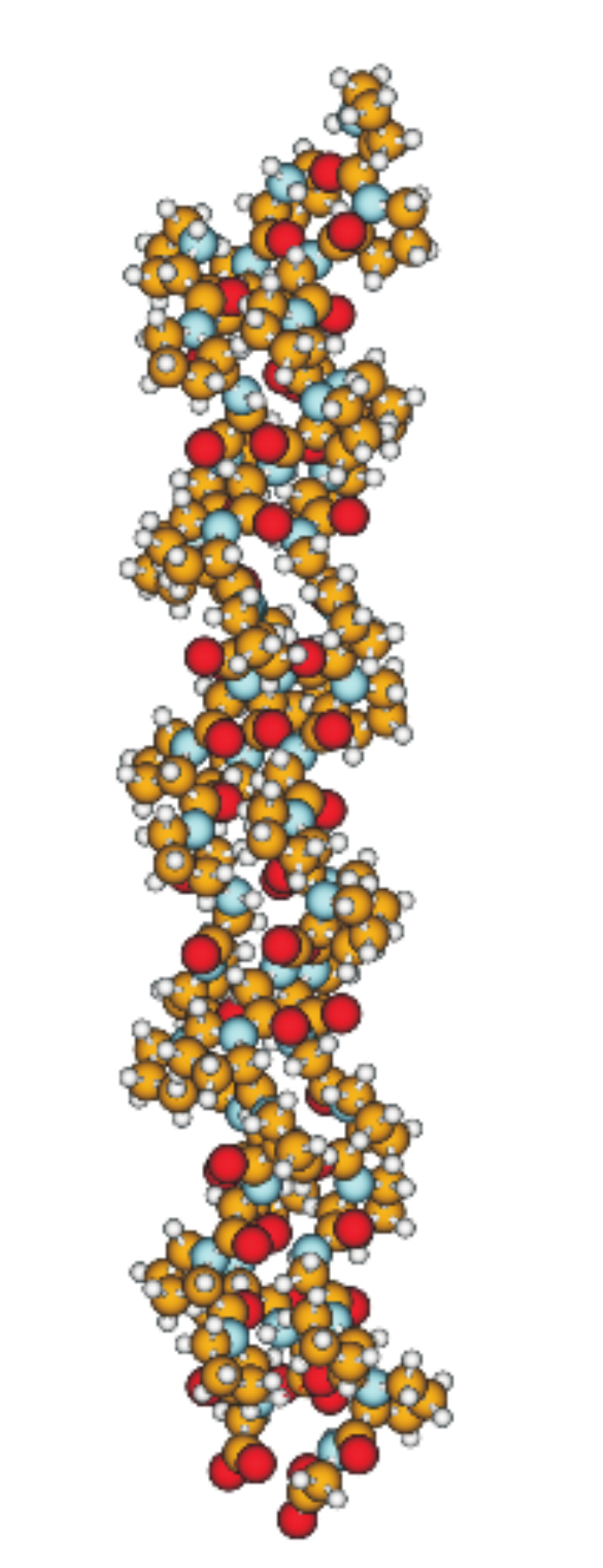}}
\label{fig:stacksub} 
\caption{PM6/PCM CP and 7/2 structures for collagen-like peptides 3(PPG)$_6$. Panel (a) top view of CP structure, (b) top view of the 7/2 structure, (c) side view of the CP structure, and (d) side view of the 7/2 structure.}
\end{figure}

\begin{figure}
\centering
\label{CDplots}
\includegraphics[width=9cm]{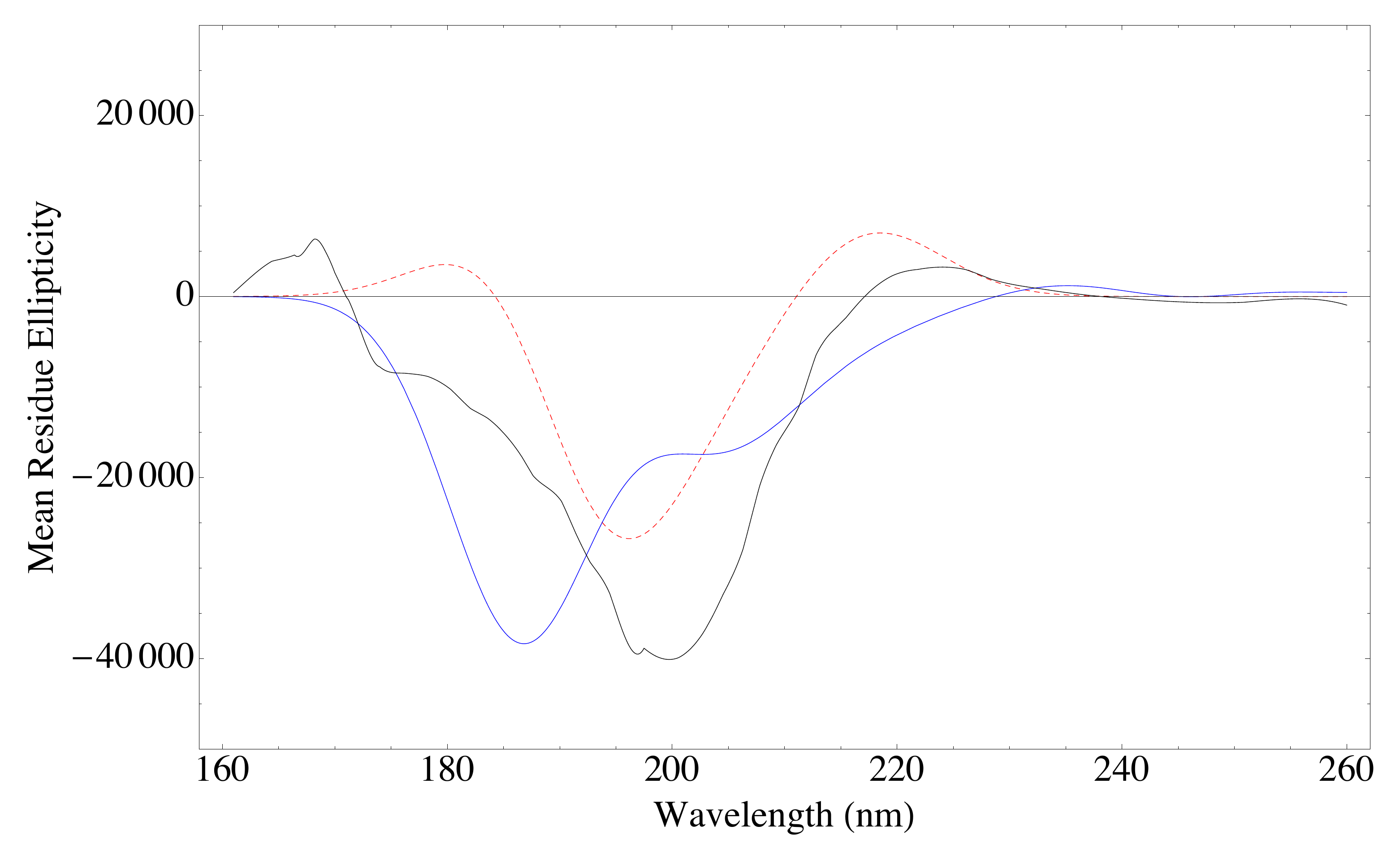}
\caption{Electronic circular dichroism  CAM-B3LYP/PCM for CP (smooth blue curve) and 7/2 (smooth dotted red curve) structures and ECD M06/PCM for CP  structure of 3(PPG)$_3$. The experimental collagen data (black curve) are replotted from Ref. \cite{miles2006} after digitizing. Mean residue ellipticity (MRE) is measured in units of degrees$\cdot$cm$^2\cdot$dmol$^{-1}$.}
\end{figure}

\begin{figure}
\centering
\subfloat[]{
\label{fig:stacksub:a} 
\includegraphics[width=0.25\linewidth]{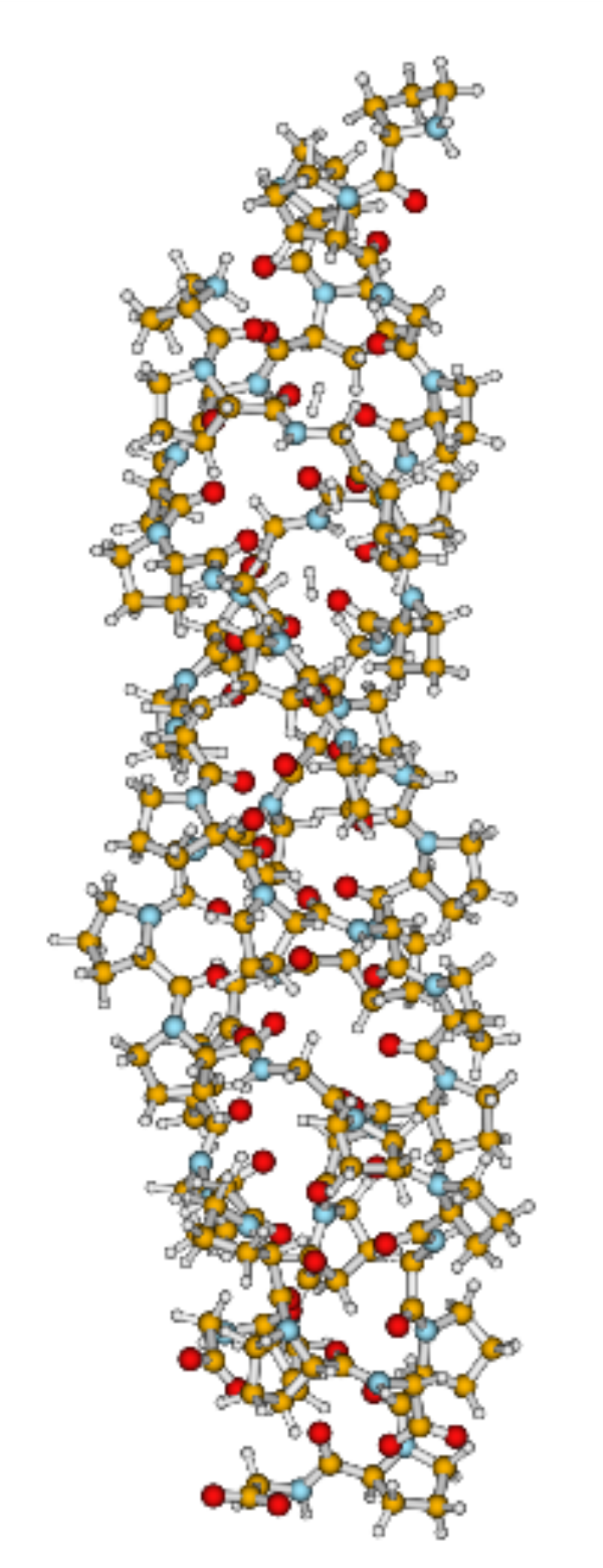}}
\hspace{0.1\linewidth}
\subfloat[]{
\label{fig:stacksub:b} 
\includegraphics[width=0.50\linewidth]{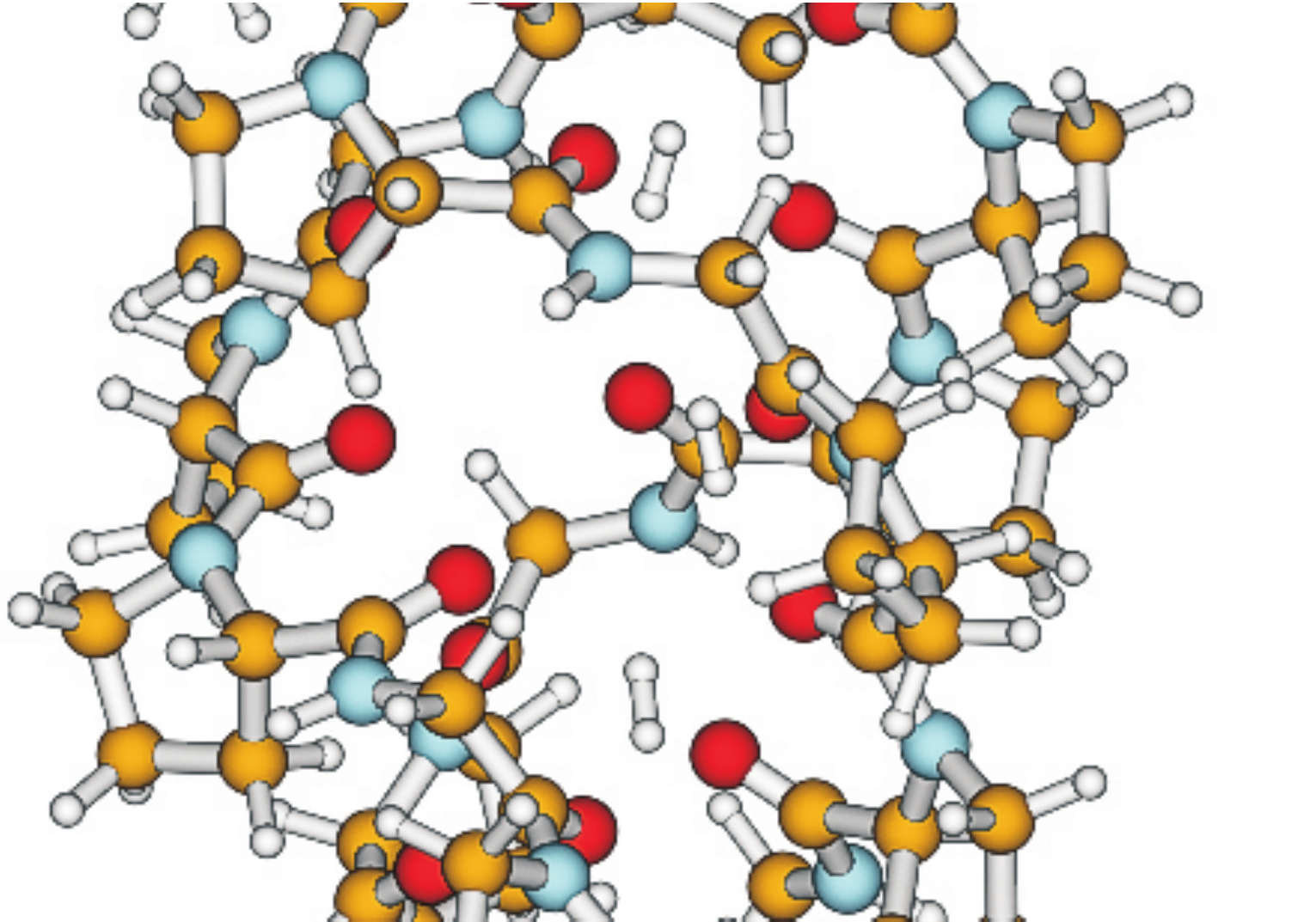}
}\\[20pt]
\label{fig:stacksub} 
\caption{(a) A stick and ball model of the close-packed 3(PGG)$_6$ structure with four H$_2$ molecules in the central channel. (b) Expanded view of the part of the structure containing the H$_2$ molecules. The vertical distance between individual H$_2$ molecules is about 6 \AA.}
\end{figure}

In Fig. 1 the PM6/PCM CP and 7/2 structures for 3(PGG)$_6$, top and side views are depicted for both systems. As one can see in Fig. 1(a), there is a visible open channel in the CP structure surrounded by oxygen atoms. In  Tables  \ref{tab:1} and \ref{tab:2} we present  the backbone and side chain torsional angles for all of the residues for the three chains for the CP structures 3(PGG)$_3 $ and 3(PGG)$_6 $. In the Supplementary Material we present the backbone and side chain angles for the complete 3(PGG)$_6 $ CP structure as well as  its corresponding xyz file.

In our analysis, the CP and 7/2 structures have nearly the same energy. The PM6 energy difference between the 7/2 structure and CP structure for collagen is $E(7/2)-E(CP)=-1.739$ eV, or $-0.065$ eV per residue for 3(PGG)$_3$, and $E(7/2)-E(CP)=-29.58$ eV, or $-0.37$ eV per residue for 3(PGG)$_6$. This higher energy difference for the longer structures reflects the difficulties that the applied methods have in properly optimizing long helical structures.

\begin{table}
\caption{\label{tab:1}PM6/PCM 3[PPG]$_3$ CP structure: backbone and side chain angles (deg)}
\begin{center}
\begin{tabular}{lrrr}
\hline\noalign{\smallskip}
Torsion angles                                                                        & chain 1     & chain 2 & chain 3 \\
\hline\noalign{\smallskip}
$\phi$'(C$_\delta$NC$_\alpha$C')                                        &   -62   &   -60  &  -63  \\
$\psi_1$(NC$_\alpha$C'N)                                                    &   136   &   141  &  150  \\
$\omega_1$(C$_\alpha$C'NC$_\alpha$)                              &  -176   &   177 &    169  \\
$\xi_{1,1}$(NC$_\alpha$C$_\beta$C$_\gamma$)               &     -7  &      -14  &   -15  \\
$\xi_{1,2}$(C$_\alpha$C$_\beta$C$_\gamma$C$_\delta$) &   14   &     19  &  22  \\
$\xi_{1,3}$(C$_\beta$C$_\gamma$C$_\delta$N)                &     -16  &      -17  &   -20  \\
$\xi_{1,4}$(C$_\gamma$C$_\delta$NC$_\alpha$)               &   12  &     9  &  11  \\
$\phi_2$(C'NC$_\alpha$C')                                                    &     -73  &      -71 &      -92  \\
$\psi_2$(NC$_\alpha$C'N)                                                     &   148  &   141  &   151  \\
$\omega_2$(C$_\alpha$C'NC$_\alpha$)                               &  172  & 179 &  -176  \\
$\xi_{2,1}$(NC$_\alpha$C$_\beta$C$_\gamma$)                &    14  & 14  &   21  \\
$\xi_{2,2}$(C$_\alpha$C$_\beta$C$_\gamma$C$_\delta$) &     -14  &  -11 &    -18  \\
$\xi_{2,3}$(C$_\beta$C$_\gamma$C$_\delta$N)                 &       10  &   5  &   8  \\
$\xi_{2,4}$(C$_\gamma$C$_\delta$NC$_\alpha$)               &     -1  & 4  &  6  \\
$\phi_3$(C'NC$_ \alpha$C')                                                   &     -76  &  -95  &  -117  \\
$\psi_{3,1}$(NC$_\alpha$C'O1)                                            &   -59  & 109  & 58  \\
$\psi_{3,1}$(NC$_\alpha$C'O2)                                            &       169  &    179  &   173  \\
\hline\noalign{\smallskip}
\end{tabular}
\end{center}
\end{table}

\begin{table}
\caption{\label{tab:2}PM6/PCM 3[PPG]$_6$ CP structure: Average backbone and side chain angles (deg) for the two middle PPG groups.}
\begin{center}
\begin{tabular}{lrrr}
\hline\noalign{\smallskip}
Torsion angles                                                                        & chain 1     & chain 2 & chain 3 \\
\hline\noalign{\smallskip}
$\phi$'(C$_\delta$NC$_\alpha$C')                                        &   -21   &   -23  &  -27  \\
$\psi_1$(NC$_\alpha$C'N)                                                    &   176   &   180  &  178  \\
$\omega_1$(C$_\alpha$C'NC$_\alpha$)                              &  -176   &   177 &    177  \\
$\xi_{1,1}$(NC$_\alpha$C$_\beta$C$_\gamma$)               &     10  &      6  &   12  \\
$\xi_{1,2}$(C$_\alpha$C$_\beta$C$_\gamma$C$_\delta$) &   -32   &     -28  &  -33  \\
$\xi_{1,3}$(C$_\beta$C$_\gamma$C$_\delta$N)                &     41  &      40  &   41  \\
$\xi_{1,4}$(C$_\gamma$C$_\delta$NC$_\alpha$)               &   -36  &     -38  &  -35  \\
$\phi_2$(C'NC$_\alpha$C')                                                    &     12  &      7  &      13  \\
$\psi_2$(NC$_\alpha$C'N)                                                     &   146  &   141  &   150  \\
$\omega_2$(C$_\alpha$C'NC$_\alpha$)                               &  -176  & -172 &  -179  \\
$\xi_{2,1}$(NC$_\alpha$C$_\beta$C$_\gamma$)                &    -44  & -48  &   -45  \\
$\xi_{2,2}$(C$_\alpha$C$_\beta$C$_\gamma$C$_\delta$) &     27  &  36  &    29  \\
$\xi_{2,3}$(C$_\beta$C$_\gamma$C$_\delta$N)                 &       0  &   -11  &   -3  \\
$\xi_{2,4}$(C$_\gamma$C$_\delta$NC$_\alpha$)               &     -27  & -18  &  --25  \\
$\phi_3$(C'NC$_ \alpha$C')                                                   &     -94  &  -75  &  -93  \\
$\psi_{3,1}$(NC$_\alpha$C'O1)                                            &   -130  & -124  & -126  \\
$\psi_{3,1}$(NC$_\alpha$C'O2)                                            &       -170  &    -176  &    -176  \\
\hline\noalign{\smallskip}
\end{tabular}
\end{center}
\end{table}

Collagen has also been studied by measurements of circular dichroism (CD) spectra \cite{bhatnagar1996,wallace2001,miles2006}. Figure 2 shows the calculated ECD spectra of the CP and 7/2 motifs along with experimental data. The experimental ECD spectra is reproduced from Ref. \cite{miles2006}. The data shows a characteristic negative peak at around 200 nm and a smaller positive peak at 220 nm. For the CAM-B3LYP/KS-TD-DFT/PCM electronic CD (EDC) calculated spectra for the PM6/PCM 3(PGG)$_3$ structures the semi-empirical PM6 WFT was used for the geometry optimizations and KS-TD-DFT with the CAM-B3LYP exchange correlation (XC) functional was used for the ECD calculations. 
The EA and ECD spectra were all simulated with FWHM line widths of 15 nm. 
As one can see by comparison, neither of the two calculations describe the spectra significantly better than the other. The 7/2 model is somewhat better at predicting the  positive peak around 220 nm. We have observed that the detailed shape of the calculated spectra is sensitive to the exact values of the refined coordinates and hence we must assume some level of uncertainty simply due to the accuracy of the methods. The integrated intensities of the spectra (Ref. \cite{miles2006}), of the CP, and of the 7/2 are $-8.4 \times 10^{-2}$, $-9.2 \times 10^{-2}$, and $-2.9 \times 10^{-2}$ degrees$\cdot$cm$^3$$\cdot$dmol$^{-1}$, respectively.  Hence, the CP calculation is more in agreement with regards to the negative bias. The reason for this is that the CP structure does not have the right-handed super coil geometry that reduces the overall left-handed chirality of the 7/2 structure.

Surprisingly, the central channel of the CP structure seen in Fig. 1(a) opens the possibility for collagen to hold, intercalate, and transport small molecules. In Ref. \cite{bohr2011}, the diameter of the channel was estimated to be about 2 \AA. In Fig. 3 we depict the minimization of the 3(PPG)$_6$ structure with four H$_2$ molecules in the channel. First we placed eight atomic hydrogens in the channel with an interspacing distance of about 3.5 \AA, the purpose was to see if the hydrogens would interact with the oxygens of the collagen strands, or if they would interact with each other to form molecular hydrogen? The result was that molecular hydrogens were formed. It is a demonstration of the relative stability of the CP structure that it is adaptable to such changes. 

In this Letter we have reported the fascinating result that triple helix structures for collagen, which fulfill the close-packing principle, are atomically possible and chemically plausible according to the performed minimization. Whether this is a physical result or an effect of the applied minimization algorithms remains to be seen. First principles semi-empirical WFT and KS-DFT calculations of the structures and properties of collagen and compared the close-packed motif (CP) with the supercoiled motif (7/2) have been performed. It is shown that within the method applied both structures behave as stable structures, and both have roughly the same energy. Explicit solvent molecules can not only change the relative energy of various conformations peptides, but they can even stabilize conformers and species that are not stable either in the gas phase or using the simple continuum solvent models like PCM \cite{jalkanen2008}. Complete features of the ECD spectra could not be described by neither of the two calculations while the negative bias is well described by the CP structure. Atomic hydrogens placed in the channel combine to form molecular hydrogen, and may perhaps contribute to enhance the stabilization of the close-packed structure. The CP structure with its central cavity appears to be a possible hydrogen molecule channel protein as atomic hydrogen can form stable molecular hydrogen in this channel. It is an intriguing possibility that such H$_2$ molecules will have a biological significance and function.

\begin{acknowledgments}
This work is supported by the Villum Foundation. KJJ wishes to thank DKFZ and DTU for hospitality.
\end{acknowledgments}

\newpage


\begin{table}\scriptsize
\caption{Supplementary Material A: PM6/PCM 3(PPG)$_6$ CP structure: backbone and side chain angles (deg)}
\begin{center}
\begin{tabular}{l|rrr||l|rrr||l|rrr}
\hline\noalign{\smallskip}
Torsion angles                               & chain 1     & chain 2 & chain 3 & Torsion angles                          & chain 1     & chain 2 & chain 3 & Torsion angles                               & chain 1     & chain 2 & chain 3
 \\
\hline\noalign{\smallskip}
Pro$_1$                                      &           &           &         &    Pro$_7$                                 &           &           &           &           Pro$_{13}$                          &            &           &          \\ 
$\tau$(H1$^{+}$NC$_\alpha$C')                &  -102.20  &  -79.80   & -108.81 &                                              &           &           &           &                                              &            &           &          \\
$\tau$(H2$^{+}$NC$_\alpha$C')                &    16.15  &   39.02   &    9.55 &                                              &           &           &           &                                              &            &           &          \\
$\tau$(C$_\delta$NC$_\alpha$C')               &   134.31  &  156.42   &  128.63 & $\phi_7$(C'NC$_\alpha$C')                      &   -23.87  &  -28.63   &   -20.98  &      $\phi_{13}$(C$_\delta$NC$_\alpha$C')        &   -38.34   &  -15.81   &  -20.26  \\
$\psi_1$(NC$_\alpha$C'N)                     &  -165.33  &  169.54   &  168.12 & $\psi_7$(NC$_\alpha$C'N)                       &   171.73  & -170.97   &   170.33  &        $\psi_{13}$(NC$_\alpha$C'N)             &   172.83   &   176.73  &  177.29  \\
$\omega_1$(C$_\alpha$C'NC$_\alpha$)            &  166.95  &   172.73  &   171.74 & $\omega_7$(C$_\alpha$C'NC$_\alpha$)             &  -178.22  & -170.14   &  -178.33  &    $\omega_{13}$(C$_\alpha$C'NC$_\alpha$)        &  -171.90   &  -171.79  & -172.24  \\
\hline
$\xi_{1,1}$(NC$_\alpha$C$_\beta$C$_\gamma$)       &  19.51   &   -11.33 &    21.01 & $\xi_{7,1}$(NC$_\alpha$C$_\beta$C$_\gamma$)        &   12.22  &   14.87   &   6.60    & $\xi_{13,1}$(NC$_\alpha$C$_\beta$C$_\gamma$)       &   15.47    &    7.64   &    6.63   \\
$\xi_{1,2}$(C$_\alpha$C$_\beta$C$_\gamma$C$_\delta$) & -36.67  &    -15.24 &   -41.10 & $\xi_{7,2}$(C$_\alpha$C$_\beta$C$_\gamma$C$_\delta$) &  -33.72  &  -35.79   & -29.24    & $\xi_{13,2}$(C$_\alpha$C$_\beta$C$_\gamma$C$_\delta$) &  -35.25    &  -30.31   &  -28.29  \\
$\xi_{1,3}$(C$_\beta$C$_\gamma$C$_\delta$N)       &  38.64   &    35.94  &    44.98 & $\xi_{7,3}$(C$_\beta$C$_\gamma$C$_\delta$N)       &   42.07  &   42.65   &  40.79    & $\xi_{13,3}$(C$_\beta$C$_\gamma$C$_\delta$N)        &   41.38    &   41.59   &   39.15  \\
$\xi_{1,4}$(C$_\gamma$C$_\delta$NC$_\alpha$)      & -27.16   &   -43.32  &   -32,20 & $\xi_{7,4}$(C$_\gamma$C$_\delta$NC$_\alpha$)       &  -36.33  &  -35.26   & -38.38    & $\xi_{13,4}$(C$_\gamma$C$_\delta$NC$_\alpha$)       &  -33.30    &  -38.43   &  -36.69  \\
\hline
Pro$_2$                                      &          &           &          &    Pro$_8$                                   &          &           &          & Pro$_{14}$                                     &             &           &          \\   
$\phi_2$(C'NC$_\alpha$C')                     &    3.52  &   -10.84  &    -8.64 & $\phi_8$(C'NC$_\alpha$C')                      &    5.59  &    1.91   &   12.53  &  $\phi_{14}$(C'NC$_\alpha$C')                    &    -9.84   &     6.93  &    5.04   \\
$\psi_2$(NC$_\alpha$C'N)                      &  151.50  &   145.37  &   126.67 & $\psi_8$(NC$_\alpha$C'N)                       &  154.23  &  146.41   &  153.71  & $\psi_{14}$(NC$_\alpha$C'N)                      &   151.61   &   139.91  &  142.65   \\
$\omega_2$(C$_\alpha$C'NC$_\alpha$)            & -167.34   &  -153.78  & -168.18  & $\omega_8$(C$_\alpha$C'NC$_\alpha$)             & -176.20  &  172.85   &  179.19  & $\omega_{14}$(C$_\alpha$C'NC$_\alpha$)            &   -178.61  &   -175.10  &  172.46    \\
\hline
$\xi_{2,1}$(NC$_\alpha$C$_\beta$C$_\gamma$)       &  -47.54   &  -36.19  &   -29.67  & $\xi_{8,1}$(NC$_\alpha$C$_\beta$C$_\gamma$)       &  -45.77  &  -48.13   &  -43.94   & $\xi_{14,1}$(NC$_\alpha$C$_\beta$C$_\gamma$)       &   -42.93   &  -44.10   &  -48.58   \\
$\xi_{2,2}$(C$_\alpha$C$_\beta$C$_\gamma$C$_\delta$)&  40.41    &  26.71   &   13.83   & $\xi_{8,2}$(C$_\alpha$C$_\beta$C$_\gamma$C$_\delta$)&  32.92  &   40.85   &   27.95   & $\xi_{14,2}$(C$_\alpha$C$_\beta$C$_\gamma$C$_\delta$)&    38.22  &    31.49  &   42.13   \\
$\xi_{2,3}$(C$_\beta$C$_\gamma$C$_\delta$N)       &  -18.32    &  -7.10  &    7.11   &  $\xi_{8,3}$(C$_\beta$C$_\gamma$C$_\delta$N)       &  -8.89  &  -19.92   &   -2.64   & $\xi_{14,3}$(C$_\beta$C$_\gamma$C$_\delta$N)       &    -20.43  &   -8.56   &  -21.55   \\
$\xi_{2,4}$(C$_\gamma$C$_\delta$NC$_\alpha$)      &  -11.59    &  -15.98  &   -26.25  &  $\xi_{8,4}$(C$_\gamma$C$_\delta$NC$_\alpha$)     & -19.69   & -10.12    & -24.67    & $\xi_{14,4}$(C$_\gamma$C$_\delta$NC$_\alpha$)      &     -6.37  &  -18.71   &  -8.71    \\
\hline
Gly$_3$                                      &           &           &          &    Gly$_9$                                    &           &          &            & Gly$_{15}$                                  &            &           &           \\
$\phi_3$(C'NC$_ \alpha$C')                    &   -96.50   &  -95.00  &  -90.36   &  $\phi_9$(C'NC$_ \alpha$C')                    &  -99.80   &  -75.67  &  -86.24    & $\phi_{15}$(C'NC$_ \alpha$C')                &   -71.93    &   -91.90  &  -87.76   \\
$\psi_3$(NC$_\alpha$C'N)                      &  -123.41  & -131.84   &  -134.10  &  $\psi_9$(NC$_\alpha$C'O1)                     & -120.41   & -120.49  & -135.51    & $\psi_{15}$(NC$_\alpha$C'O1)                 &  -124.14    &  -141.69  & -127.43   \\
$\omega_3$(C$_\alpha$C'NC$_\alpha$)             & -175.02   & -179.96   & -175.81   & $\omega_9$(C$_\alpha$C'NC$_\alpha$)             &  174.76   & -171.84  & -170.89   &  $\omega_{15}$(C$_\alpha$C'NC$_\alpha$)         & -179.02     & -178.55   &  175.58    \\
\hline\hline
Pro$_4$                                      &           &           &          &    Pro$_{10}$                                  &           &           &            &  Pro$_{16}$                                 &             &           &          \\ 
$\phi_4$(C$_\delta$NC$_\alpha$C')               & -26.40    &  -27.83   &   -19.11 & $\phi_{10}$(C'NC$_\alpha$C')                     & -18.85    &   -18.14  &  -32.86   & $\phi_{16}$'(C$_\delta$NC$_\alpha$C')           &   -16.67    &   -35.40  &  -17.54  \\
$\psi_4$(NC$_\alpha$C'N)                       & 174.92    &  175.71   &  173.02  & $\psi_{10}$(NC$_\alpha$C'N)                      & -179.25   &  171.95   & -174.97    & $\psi_{16}$(NC$_\alpha$C'N)                  &  -174.59    &  -164.62  &  174.74  \\
$\omega_4$(C$_\alpha$C'NC$_\alpha$)             & 179.99     & 160.11   &  151.71   & $\omega_{10}$(C$_\alpha$C'NC$_\alpha$)            & -173.67   &  163.73  &   174.71    & $\omega_{16}$(C$_\alpha$C'NC$_\alpha$)         & -160.20     & -154.66   & -166.38   \\
\hline
$\xi_{4,1}$(NC$_\alpha$C$_\beta$C$_\gamma$)       &    9.90    &   1.50    &   -23.20 & $\xi_{10,1}$(NC$_\alpha$C$_\beta$C$_\gamma$)       &   8.49    &  -2.46   &  17.61      & $\xi_{16,1}$(NC$_\alpha$C$_\beta$C$_\gamma$)      &    8.94     &   20.80   &   11.72   \\
$\xi_{4,2}$(C$_\alpha$C$_\beta$C$_\gamma$C$_\delta$) & -31.60    &   23.95   &     2.67 & $\xi_{10,2}$(C$_\alpha$C$_\beta$C$_\gamma$C$_\delta$) & -29.91   &  -21.03   & -36.90     & $\xi_{16,2}$(C$_\alpha$C$_\beta$C$_\gamma$C$_\delta$) &  -29.67     & -38.56    & -33.80   \\
$\xi_{4,3}$(C$_\beta$C$_\gamma$C$_\delta$N)       &   41.15    &  37.37    &   18.97  & $\xi_{10,3}$(C$_\beta$C$_\gamma$C$_\delta$N)       &  39.72   &   36.68   &  41.40     & $\xi_{16,3}$(C$_\beta$C$_\gamma$C$_\delta$N)       &   39.10     &   41.49   &  43.06   \\
$\xi_{4,4}$(C$_\gamma$C$_\delta$NC$_\alpha$)      &   -36.76   &  -38.23   &  -34.40  & $\xi_{10,4}$(C$_\gamma$C$_\delta$NC$_\alpha$)      &  -36.38  &  -39.80   &  -32.39     & $\xi_{16,4}$(C$_\gamma$C$_\delta$NC$_\alpha$)      &  -34.95     &  -29.84   & -37.43   \\
\hline
Pro$_5$                                      &           &           &          &    Pro$_{10}$                                 &           &           &            &  Pro$_{17}$                                    &             &            &          \\   
$\phi_5$(C'NC$_\alpha$C')                     &  -5.91     &   22.34   &   15.11  &  $\phi_{10}$(C'NC$_\alpha$C')                   &   18.57   &   12.69   &  12.57    &  $\phi_{17}$(C'NC$_\alpha$C')                    &   11.31    &   17.88    &   12.28   \\
$\psi_5$(NC$_\alpha$C'N)                      & 155.36     &  148.26   &  139.95  & $\psi_{10}$(NC$_\alpha$C'N)                     &  138.68   &  135.57   & 146.41    &  $\psi_{17}$(NC$_\alpha$C'N)                     &  143.24    &  138.64    &  137.55   \\
$\omega_5$(C$_\alpha$C'NC$_\alpha$)             & 178.86    &   178.26  &   179.68  & $\omega_{10}$(C$_\alpha$C'NC$_\alpha$)           & -176.07   &  171.00   &-177.61     & $\omega_{17}$(C$_\alpha$C'NC$_\alpha$)            &  171.00    &  172.50    & -177.94  \\
\hline
$\xi_{5,1}$(NC$_\alpha$C$_\beta$C$_\gamma$)       &  -48.80   & -43.13    &  -45.73   & $\xi_{10,1}$(NC$_\alpha$C$_\beta$C$_\gamma$)       &  -42.22   &   -48.19 &  -46.91    & $\xi_{17,1}$(NC$_\alpha$C$_\beta$C$_\gamma$)        &   -40.07   &   -42.40  & -45.58    \\
$\xi_{5,2}$(C$_\alpha$C$_\beta$C$_\gamma$C$_\delta$)&  44.97    &  20.51    &   27.71   & $\xi_{10,2}$(C$_\alpha$C$_\beta$C$_\gamma$C$_\delta$)&  20.35   &    31.03  &  29.91    & $\xi_{17,2}$(C$_\alpha$C$_\beta$C$_\gamma$C$_\delta$) &    28.09   &   27.34   &  31.62    \\
$\xi_{5,3}$(C$_\beta$C$_\gamma$C$_\delta$N)       &  -25.77    &   9.32   &    -0.10  &  $\xi_{10,3}$(C$_\beta$C$_\gamma$C$_\delta$N)       &   8.19   &    -3.29 &   -3.04    & $\xi_{17,3}$(C$_\beta$C$_\gamma$C$_\delta$N)        &    -6.91   &    -3.47  &   -7.57    \\
$\xi_{5,4}$(C$_\gamma$C$_\delta$NC$_\alpha$)      &   -4.44    &  -36.41  &   -28.61  &  $\xi_{10,4}$(C$_\gamma$C$_\delta$NC$_\alpha$)       & -33.90   &   -26.80 &  -25.99    & $\xi_{17,4}$(C$_\gamma$C$_\delta$NC$_\alpha$)       &   -17.93   &   -22.79  &  -20.46    \\
\hline
Gly$_6$                                      &           &           &          &    Gly$_{12}$                                   &           &           &           & Gly$_{18}$                                     &            &           &           \\
$\phi_6$(C'NC$_ \alpha$C')                    &   -82.75   &  -88.83  &   -84.60  &  $\phi_{12}$(C'NC$_ \alpha$C')                    &  -88.71  &  -74.84  & -100.12   & $\phi_{18}$(C'NC$_ \alpha$C')                     &    -70.80  &   -98.45  &   -89.04  \\
$\psi_6$(NC$_\alpha$C'N)                      &  -127.38   & -136.72  &  -123.46  &  $\psi_{12}$(NC$_\alpha$C'O1)                     & -138.87  & -127.36  & -115.61   & $\psi_{18}$(NC$_\alpha$C'O1)                      &      7.74  &    38.56  &    24.69  \\
$\omega_6$(C$_\alpha$C'NC$_\alpha$)             &  -171.50  &  -171.41  &  -179.43  & $\omega_{12}$(C$_\alpha$C'NC$_\alpha$)             & -165.94  &  179.81  &  178.98   & $\psi_{18}$(NC$_\alpha$C'O2)                     &   -172.34  &  -141.93  &  -156.42   \\
\hline\noalign{\smallskip}
\end{tabular}
\end{center}
\end{table}

\end{document}